\documentclass[twocolumn]{aastex631}

\usepackage{graphicx}
\usepackage{amsmath}
\usepackage{color}
\usepackage{ulem}
\usepackage{float}
\usepackage{soul}
\graphicspath{{fig/}}

\newcommand\hcnone{HCN $J=1 \rightarrow 0$}
\newcommand\hcopone{HCO$^+\ J=1 \rightarrow 0$}
\newcommand\coone{CO $J=1 \rightarrow 0$}
\newcommand\cotwo{CO $J=2 \rightarrow 1$}

\newcommand\lx{$L_{\rm 0.5 - 2\,keV}^{\rm gas}$}

\newcommand\hcop{HCO$^+$}

\begin{document}

\title{Fire and Ice in the Whirlpool: Spatially Resolved Scaling Relations \\ between X-ray Emitting Hot Gas and Cold Molecular Gas in M51}
\correspondingauthor{Junfeng Wang}
\email{jfwang@xmu.edu.cn}

\author{Chunyi Zhang}
\affiliation{Department of Astronomy, Xiamen University, 422 Siming South Road, Xiamen 361005, People's Republic of China}

\author[0000-0003-4874-0369]{Junfeng Wang}
\affiliation{Department of Astronomy, Xiamen University, 422 Siming South Road, Xiamen 361005, People's Republic of China}

\author{Tian-Wen Cao}
\affiliation{Department of Astronomy, Xiamen University, 422 Siming South Road, Xiamen 361005, People's Republic of China}


\begin{abstract}

The cold and hot interstellar medium (ISM) in star forming galaxies resembles the reservoir for star formation and associated heating by stellar winds and explosions during stellar evolution, respectively. We utilize data from deep $Chandra$ observations and archival millimeter surveys to study the interconnection between these two phases and the relation to star formation activities in M51 on kiloparsec scales. A sharp radial decrease is present in the hot gas surface brightness profile within the inner 2 kpc of M51. The ratio between the total infrared luminosity ($L_{\rm IR}$) and the hot gas luminosity ({\lx}) shows a positive correlation with the galactic radius in the central region. For the entire galaxy, a twofold correlation is revealed in the {\lx}${-}$$L_{\rm IR}$ diagram, where {\lx} sharply increases with $L_{\rm IR}$ in the center but varies more slowly in the disk. The best fit gives a steep relation of ${\rm log}(L_{\rm 0.5-2\,keV}^{\rm gas} /{\rm erg\,s^{-1}})=1.82\,{\rm log}(L_{\rm IR} /{L_{\rm \odot}})+22.26$ for the center of M51. The similar twofold correlations are also found in the {\lx}${-}$molecular line luminosity ($L^\prime_{\rm gas}$) relations for the four molecular emission lines CO(1-0), CO(2-1), HCN(1-0), and {\hcop}(1-0). We demonstrate that the core-collapse supernovae (SNe) are the primary source of energy for heating gas in the galactic center of M51, leading to the observed steep {\lx}${-}$$L_{\rm IR}$ and {\lx}${-}$$L^\prime_{\rm gas}$ relations, as their X-ray radiation efficiencies ($\eta$ $\equiv$ {\lx}/$\dot{E}_\mathrm{SN}$) increase with the star formation rate surface densities ($\Sigma_{\rm SFR}$), where $\dot{E}_\mathrm{SN}$ is the SN mechanical energy input rate.

\end{abstract}


\section{Introduction}\label{sec:intro}

Since the pioneering studies of \citet{Gao&Solomon2004ApJ_a,Gao&Solomon2004ApJS_b} which revealed a tight linear correlation between the luminosities of infrared emission and line emission {\hcnone} from dense gas in nearby galaxies, it has been clear that the dense molecular gas is the raw material for star formation rather than the atomic gas. The dense gas is typically defined as gas with a volume density greater than 10$^4$ cm$^{-3}$ and can be traced by molecular emission lines with high critical densities like HCN \citep{2008Baan, 2010LadaApJ, 2014Zhang, Tan2018, 2019Jimenez_empire_880_127}. Subsequent observations in the last two decades have shown that the linear correlation extends from local Galactic clouds to high-redshift galaxies and spans almost 12 orders of magnitude in infrared and HCN luminosity \citep{Wu2005ApJ, Gao2007ApJ, 2008Baan, 2010Liu&Gao_713_524, 2011wang&zhang&shi_416_L21, 2012Garcia-Burillo&Usero_539_A8, 2015Chen&Gao, 2015Usero, 2019Jimenez_empire_880_127, Jiang2020MNRAS}. 

Furthermore, it is well established that the amount of X-ray emission and far-infrared luminosity of star-forming galaxies show a linear relationship on a global scale \citep{Ranalli2003A&A}. This is well understood due to a part of UV radiation from massive and young stars embedded in dust, which is absorbed by dust grains and reradiated in the infrared band. Meanwhile, the short-lived massive stars produce substantial amounts of hot ionized gas at $\thicksim$ sub-keV temperatures, serving as a primary source of soft X-ray emission \citep{Grimes2005ApJ, 2009OwenMNRAS, 2012Mineo2, Li2013MNRAS1, 2022LehmerApJ, Kyritsis2024arXiv}. Further, the collective luminosity of high-mass X-ray binaries dominating the hard X-ray band is tightly correlated with the star formation rate (SFR) of the host galaxy as traced by near-ultraviolet and total infrared emission \citep{2012Mineo1}.

However, the linear relation between X-ray emission, SFR, and ionized or molecular interstellar gas does not appear to hold on more resolved physical scales. \citet{2020Kouroumpatzakis} divided 13 star-forming galaxies by the grids (1$\times$1, 2$\times$2, 3$\times$3, and 4$\times$4 kpc$^2$) of different sub-galactic scales and found shallower slopes of the $L_{\rm X}$${-}$SFR relations in logarithmic space for three SFR indicators (e.g., H$_\alpha$, 8$\mu$m, and 24$\mu$m). In contrast, \citet{Zhang2024ApJ} obtained a steep relation characterized by a power-law index of $n \approx 1.8$ between hot gas luminosity and SFR, as traced by the total infrared band, for the nuclear regions of five nearby star-forming galaxies on sub-kiloparsec scales ($\thicksim$ 200 pc). One possible explanation for the slope discrepancies is the difference in the physical scales of the regions studied. \citet{Zhang2024ApJ} focus on the inner $\thicksim$ 1.6 kpc regions of the five galaxies, which only correspond to merely one grid region on the sub-galactic scale. Therefore, the results of \citet{2020Kouroumpatzakis} actually show the relationships between X-ray emission and SFR for the galactic disks at different spatial scales. 

To determine whether hot gas emission and SFR have different scaling relations in the galactic center and disk, and to explore the connection between X-ray hot gas and cold gas traced by molecular emission lines, we study the nearby star-forming galaxy M51 on a scale of $\thicksim$ 1.3 kpc. In this Letter, we present the comparison between the diffuse X-ray emission and total infrared luminosity and four molecular emission lines CO(1-0), CO(2-1), HCN(1-0), and {\hcop}(1-0) across the entire galaxy.

\begin{figure}
    \includegraphics[scale=0.775]{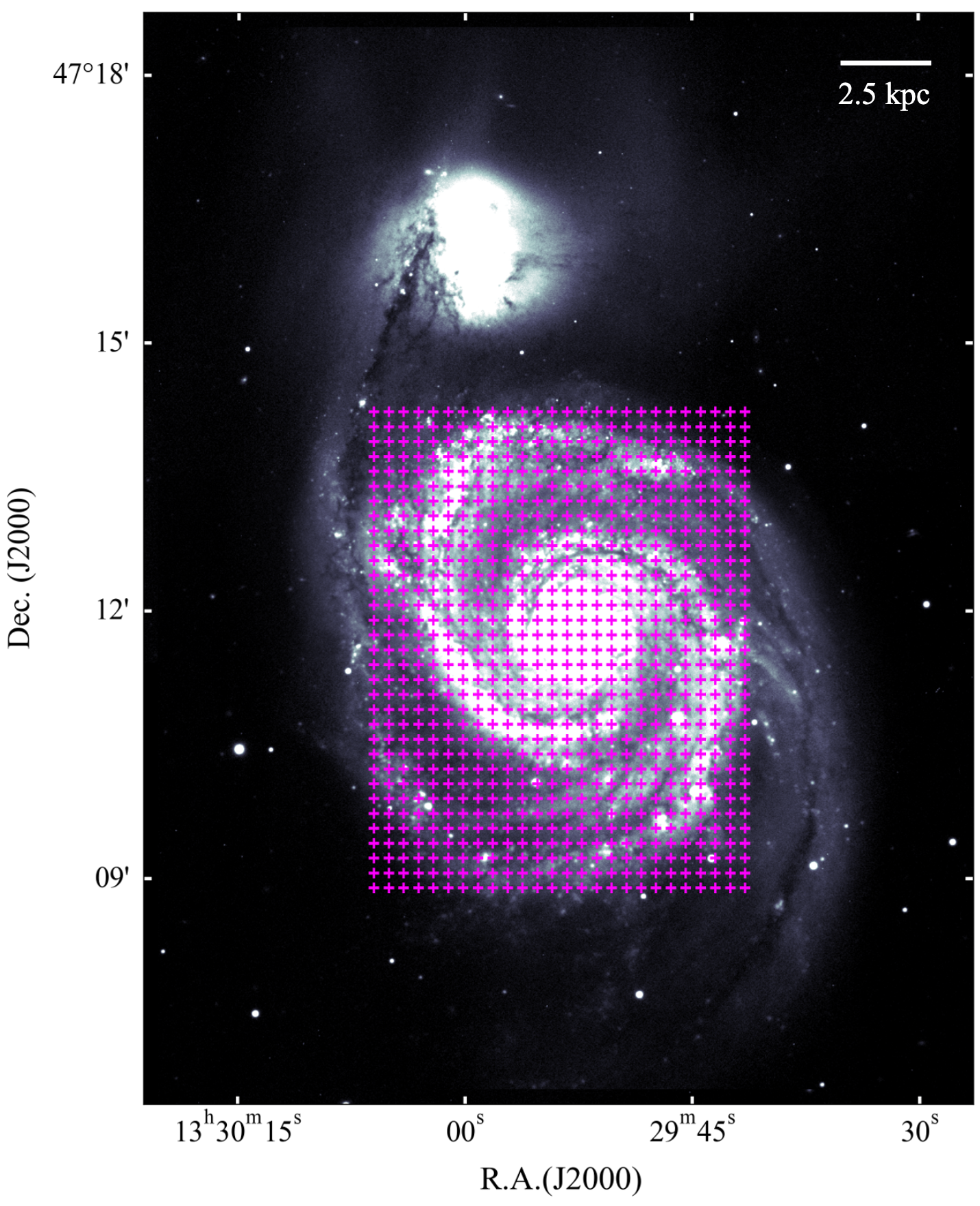}
    \caption{The central positions of regions used in this work, marked by purple crosses overlaid on the KPNO \emph{V}-band image of M51. The purple crosses are spaced by 10$\arcsec$. Each region corresponds to a circle with a diameter of 33$\arcsec$ and is aligned with the millimeter data.}
    \vspace{0.2cm}
    \label{m51}
\end{figure}

\section{Observations and data reduction} \label{sec:data}

\subsection{Millimeter data }\label{sec:millimeter}

We obtained the {\hcnone} and {\hcopone} data from the \emph{EMIR Multiline Probe of the ISM Regulating Galaxy Evolution} \citep[EMPIRE;][]{2019Jimenez_empire_880_127}, {\coone} data from the \emph{PdBI Arcsecond Whirlpool Survey} \citep[PAWS;][]{2013paws}, {\cotwo} data from the \emph{HERA CO-Line Extragalactic Survey} \citep[HERACLES;][]{2009Heracles}. All molecular emission lines were observed by the IRAM 30-m single dish telescope\footnote{The millimeter data can be found at \url{https://iram-institute.org/science-portal/proposals/lp/completed/}}. We regridded the millimeter data using a grid spacing of 10$\arcsec$ and then aligned them with the HCN data. To compare the emission lines, we convolved all data with Gaussian kernels to 33$\arcsec$ $\approx$ 1.3 kpc working resolution at the 8.4 Mpc distance to M51 \citep{Tully2009AJ}.

The molecular emission line intensities were measured as $I \tbond \int T_{\rm mb}dv$ over the emission window, where $T_{\rm mb}$ was the main beam temperature and $v$ was the velocity. We computed the uncertainties on the integrated intensity as:
\begin{equation}\label{eq1}
    \sigma=T_{\rm rms} \sqrt{\Delta v_{\rm line} \Delta v_{\rm res}} \sqrt{1+\Delta v_{\rm line}/\Delta v_{\rm base}},
    \end{equation}
where $T_{\rm rms}$ was the 1$\sigma$ RMS value of the noise for a spectrum with velocity resolution of $\Delta v_{\rm res}$, $\Delta v_{\rm line}$ was the velocity range of the emission line, and $\Delta v_{\rm base}$ was the velocity width used to fit the baseline. We required that the detected emission lines be at least three times the uncertainty. Following \citet{1997solomon}, the line luminosity $L^\prime_{\rm gas}$ was calculated via:
\begin{equation}\label{eq2}
    \begin{split}
    L^\prime_{\rm gas}= & 3.25\times10^7 \left(\frac{S\Delta v}{{\rm 1\ Jy\ km\ s^{-1}}}\right)\left(\frac{\nu_{\rm obs}}{{\rm 1\ GHz}}\right)^{-2}\\
    & \times\left(\frac{D_{\rm L}}{{\rm 1\ Mpc}}\right)^2 \left(1+z\right)^{-3}\ {\rm K\ km\ s^{-1}\ pc^2},
    \end{split}
    \end{equation}
where $S\Delta v$ was the velocity-integrated flux density, $\nu_{\rm obs}$ was the observed line frequency, and $D_{\rm L}$ was the luminosity distance.

\subsection{X-Ray data }\label{sec:chandra}

We utilized three $Chandra$ observations from 2012 (ObsIds 13812, 13813, and 13814), with corresponding exposure times of 157.5 ks, 179.2 ks, and 189.9 ks, respectively (see \citealt{2016Kuntz_ApJ} for details of the observations). The data were reprocessed with CIAO v.4.13 and CALDB v.4.9.6. To obtain the emission of hot gas, we subtracted point-like sources detected by using the wavelet-based source detection algorithm \emph{wavdetect} \citep{Freeman2002ApJS} on the scales of 1, $\sqrt{2}$, 2, 2$\sqrt{2}$, 4, 4$\sqrt{2}$, and 8 pixels ($0^{\prime \prime}.492$/pixel) and in the soft (0.5${-}$2.0 keV), hard (2.0${-}$8.0 keV) and total (0.5${-}$8.0 keV) energy bands. The 90\% energy-encircled ellipse was adopted as the source region radius. For brighter sources, we manually increased the size of the ellipses to remove visible ring-like features due to the point spread function (PSF) wing to minimize the contamination from the PSF spillover of the compact sources. 

The diffuse X-ray spectra were extracted using the \emph{specextract} task, and the background spectra were obtained from source-free regions within the same field. We combined the multiple spectra for each position with the \emph{combine$\_$spectra} task. In order to use ${\chi}^2$ as the fit statistic, the combined spectra were rebinned to achieve a minimum of 10 counts per bin. The XSPEC v.12.11.0 \citep[][]{1996ASPC..101...17A} software was used to fit the spectra, which were well represented by an absorbed thermal plasma model APEC \citep{2001ApJ...556L..91S}, supplemented with Gaussian emission-line components as necessary. We corrected all the X-ray luminosities for both Galactic and intrinsic absorption. Comparing with the observed soft band flux, we found that correction for absorption introduced a factor of $\thicksim$ 7, consistent with the test case of NGC 3256 in which only 10\% of the 0.5${-}$2.0 keV flux could penetrate through the obscuring gas \citep{Ranalli2003A&A}.

\begin{figure}
    \includegraphics[scale=0.565]{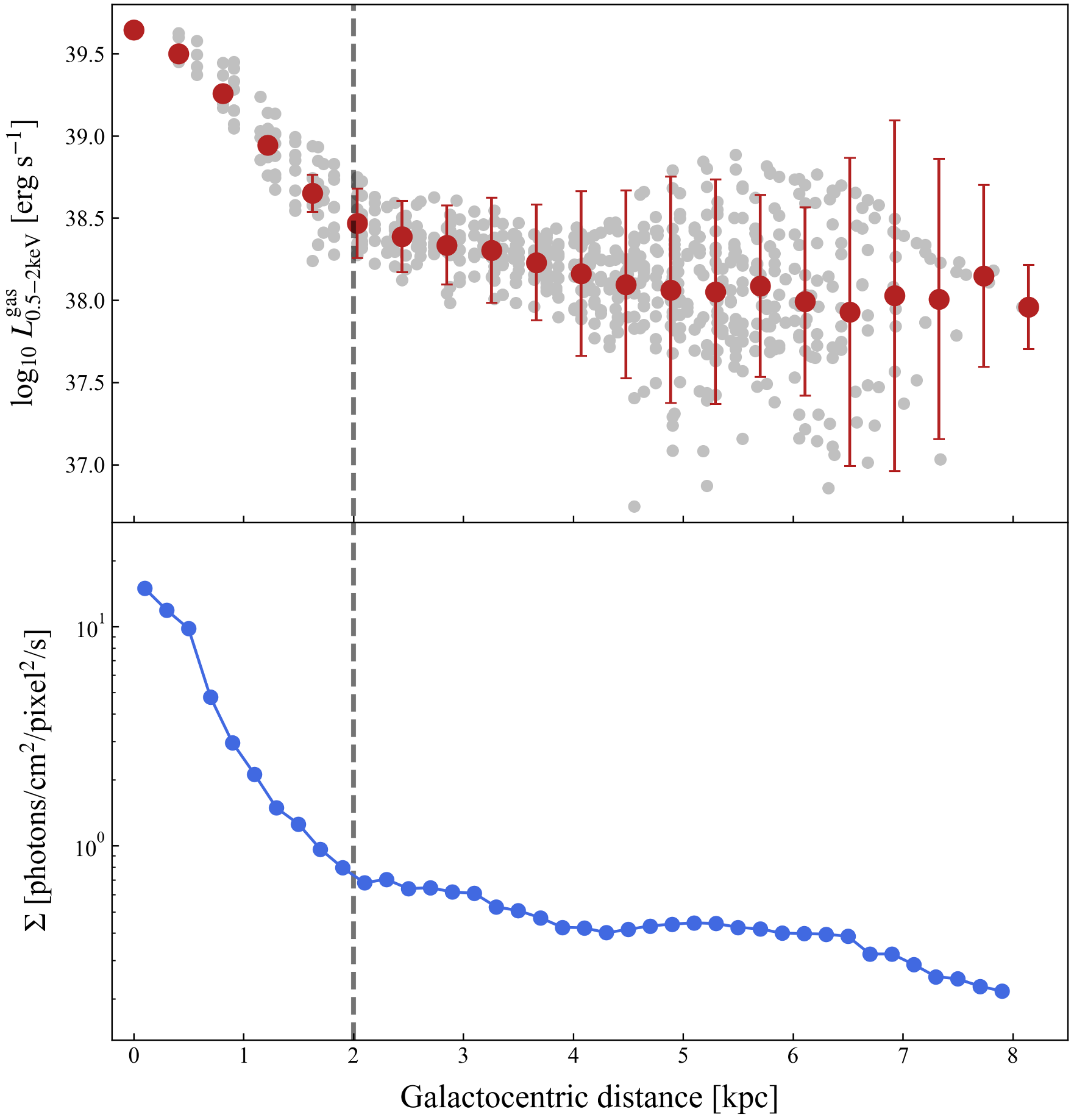}
    \vspace{-0.3cm}
    \caption{Radial distribution of hot gas luminosity and surface brightness (0.5${-}$2.0 keV). Top panel: radial profile of luminosity. The gray points denote regions matched to the millimeter data, and the red points are average values of the X-ray luminosities within the intervals \emph{r} = \emph{n} $\times$ 0.4 kpc + 0.2 kpc $\pm$ 0.2 kpc, where \emph{n} = 0,1...19. The error bars are the statistical uncertainty of the average X-ray luminosities. Bottom panel: surface brightness profile of hot gas. The surface brightness profile is constructed in concentric annuli across the entire galaxy. The dashed line shows the radius of 2 kpc (50$\arcsec$) to distinguish the galactic center and the outer disk.}
    \vspace{0.2cm}
    \label{radial}
\end{figure}

\subsection{Infrared Data}\label{sec:ancillary}

We retrieved the PACS 70 $\mu$m, 160 $\mu$m and SPIRE 250 $\mu$m infrared images from the Herschel Science Archive (HSA). We also obtained the MIPS 24 $\mu$m data from the \emph{Spitzer Infrared Nearby Galaxies Survey} \citep[SINGS;][]{2003sings}. The data were processed to level 2 for the 24 $\mu$m and 250 $\mu$m bans, and to level 2.5 for the 70 $\mu$m and 160 $\mu$m bands. To estimate the infrared luminosity, we used the convolution kernels provided by \citet{2011Aniano_Kernels} to convolve the \emph{Spitzer} and \emph{Herschel} maps, matching the 33$\arcsec$ working resolution. For the PACS data, the images were scaled by a factor of 1.133$\times$$(33/\emph{pixel size})^2$, where the \emph{pixel size} is the length of a pixel in arcseconds, to convert units from Jy to Jy beam$^{-1}$. For the MIPS data, we first converted the pixel values from MJy sr$^{-1}$ to Jy, followed by scaling the images to Jy beam$^{-1}$. 

We measured the central pixel flux at different positions from the scaled images to obtain the flux of each infrared band in units of Jy beam$^{-1}$. We calculated the total infrared luminosity $L_{\rm IR}$ from \mbox{3 $\mu$m} to \mbox{1100 $\mu$m}, following the method described by \citet{2013Galametz}, based on the combination of the convolved \mbox{24 $\mu$m}, \mbox{70 $\mu$m}, \mbox{160 $\mu$m}, and \mbox{250 $\mu$m} luminosities:
\begin{equation}
    L_{\rm IR}=\Sigma\ c_i \nu L_\nu(i)\ L_\odot,
    \end{equation}
where $\nu L_\nu(i)$ was the resolved luminosity in a given band $i$ in units of $L_\odot$ and measured as $4\pi {\rm D}^2_\mathrm{L}(\nu f_\nu)_i$, and $c_i$ were the calibration coefficients for various combinations of $Spitzer$ and $Herschel$ bands.

\begin{figure*}
    \centering 
    \includegraphics[scale=1.04]{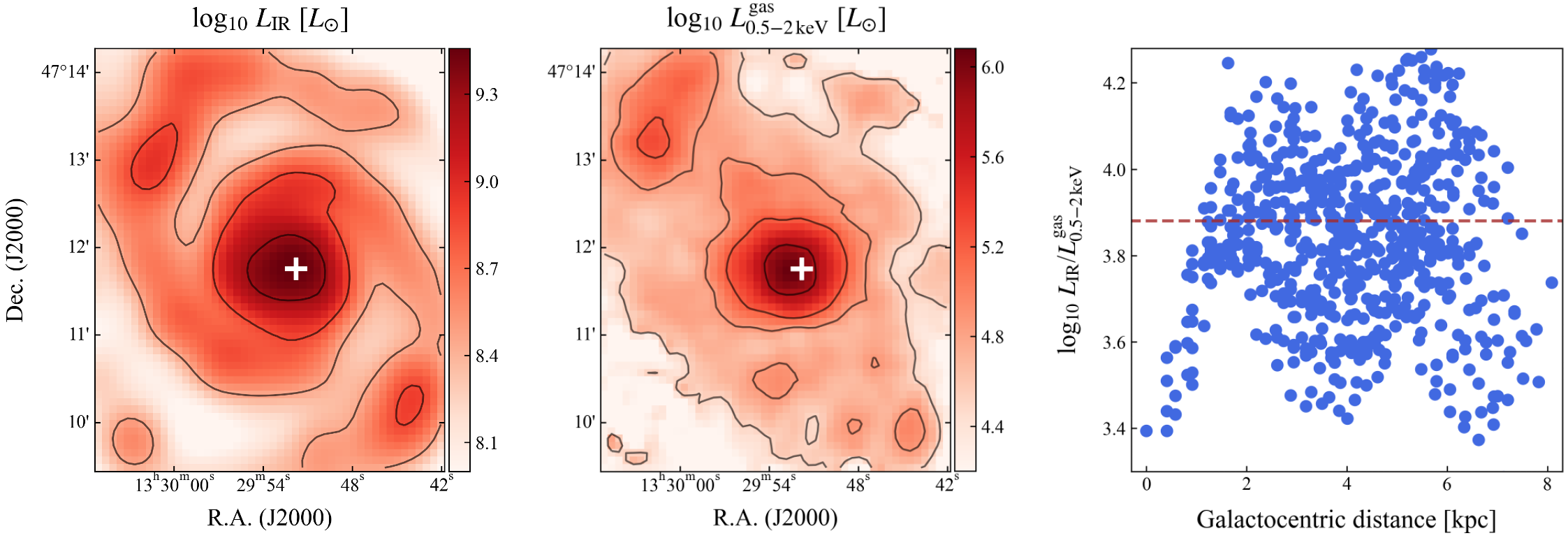}
    \vspace{-0.6cm}
    \caption{Maps of the total infrared, diffuse X-ray luminosity, and $L_{\rm IR}$/{\lx} ratios along the galactic radius. The units are given above each panel and all maps are logarithmic. Left: the infrared luminosity is convolved to 33$\arcsec$ resolution. The black lines indicate contour levels of log$L_{\rm IR}$ = 8.4, 8.9 and 9.3$L_{\rm \odot}$. Middle: the contour levels of log{\lx} = 4.5, 4.9, 5.2 and 5.8$L_{\rm \odot}$. Right: the red dashed line represents the average $L_{\rm IR}$/{\lx} ratio across the galaxy. The white crosses indicate the central position of the galaxy. }
    \vspace{0.45cm}
    \label{LTIR-Hotgas}
\end{figure*}

\section{Results}\label{sec:results}

\subsection{Hot Gas Luminosity Distribution and Surface Brightness Profile}\label{Radial profile}

To obtain the radial profile of hot gas in the 0.5$-$2 keV band, we fitted the X-ray spectra of regions with diameters of 33$\arcsec$, whose positions were aligned with the millimeter data (see Figure~\ref{m51}), and constructed a stack of concentric annuli to determine the surface brightness flux. Figure~\ref{radial} presents the distribution of luminosities and the surface brightness profile. We calculate the average luminosities over rings of 0.4 kpc width to obtain a distinct luminosity profile. It is apparent that both the luminosity and surface brightness profile decline sharply within the central 2 kpc (50$\arcsec$), followed by a more gradual variation at larger radii. This is similar to the results of CO, HCN and IR intensities of \citet{2015Chen&Gao}, who studied the spatially resolved observations of {\hcnone} in M51 and assumed a center within the central 45$\arcsec$. Following \citet{2015Chen&Gao}, we define the central 50$\arcsec$ region as the ``center'' of M51, and the region beyond this range as the ``disk'' in this work.

\subsection{Correlation between the Infrared Luminosity and the 0.5-2 keV X-Ray Luminosity}\label{Lx-LIR}

Figure~\ref{LTIR-Hotgas} shows the spatial distribution of the total infrared and hot gas luminosities, along with their ratios as a function of the galactic radius. Compared with a clear spiral structure seen in the total infrared luminosity, the emission of hot gas in the galactic disk is particularly extended. Both infrared and X-ray images show a bright core, but the ratios of $L_{\rm IR}$ to {\lx} in the center are lower than the average $L_{\rm IR}$/{\lx} ratio for the galaxy and increase along the radius. We note that the interarms have similarly low $L_{\rm IR}$/{\lx} ratios as the innermost regions of M51. The former arises from the low infrared luminosities in the interarms, and the latter implies a higher concentration of hot gas in the galactic center. A similar spatial distribution of the hot gas brightness was found by \citet{2000Pietsch}, who studied the nearby starburst galaxy NGC 253 using the deep {\em ROSAT} observations. They indicated that a third of the diffuse soft X-ray emission of the whole galaxy was in the nuclear area ($67^{\prime \prime}$.5), which accounts for $\thicksim$ 25\% of the total X-ray luminosity. 

\begin{figure}
    \includegraphics[scale=0.49]{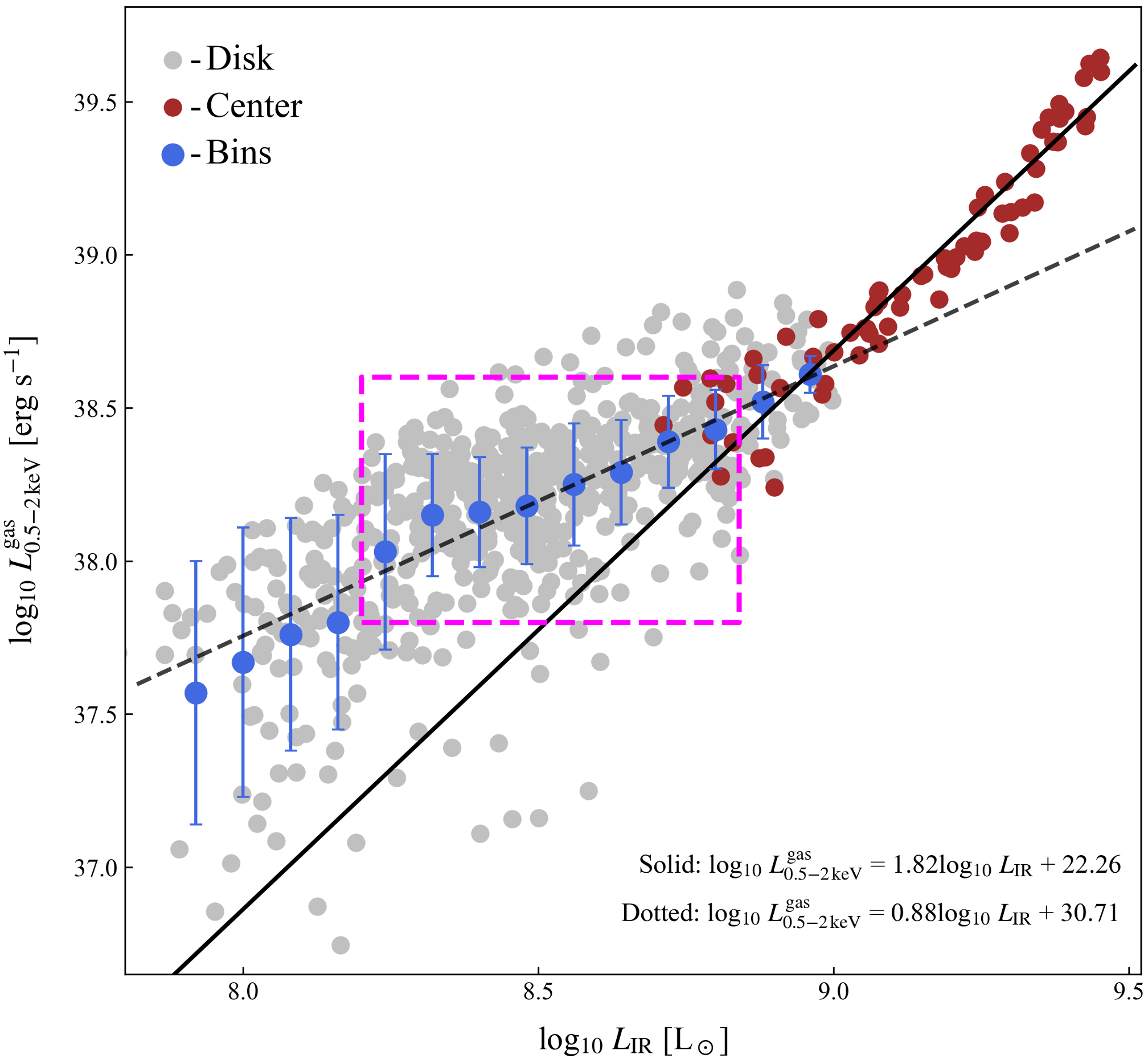}
    \vspace{-0.5cm}
    \caption{Correlations between the emission of hot gas and the total infrared luminosity across the entire galaxy at the kiloparsec scale. The red points represent regions within the central 2 kpc, and the gray points are regions located in the disk. The blue points show the binned trend for the X-ray luminosities across the disk. Each total infrared luminosity bin spacing is 0.08 dex. The solid and dashed lines are the best fit of the {\lx}${-}$$L_{\rm IR}$ relations for the center and the mean X-ray luminosities. The purple box includes almost 80\% of the disk regions, which presents a more gradual trend.}
    \vspace{0.0cm}
    \label{disk_center}
\end{figure}

To further explore the correlations between the total infrared and the hot gas luminosities across the entire galaxy at the kiloparsec scale, we plot the {\lx}${-}$$L_{\rm IR}$ relation in Figure~\ref{disk_center}. The regions within the central 2 kpc show a notably steeper trend than those in the disk, which display significant scatter in the low total infrared luminosity regime. We note that approximately 80\% of the regions in the disk are localized within the purple box, characterized by a more gradual trend. To obtain an averaged profile for the disk, we calculate the mean X-ray luminosity of the hot gas within each 0.08 dex bin of the total infrared luminosity in logarithmic space. The Bayesian linear regression method implemented in the IDL routine LINMIX\_ERR \citep{2007Kelly} was applied to the central region data and the mean X-ray luminosities. The parameters in the method were estimated using the posterior median, with the error determined as the median absolute deviation of the posterior distribution. The best-fit relations with uncertainties are listed below:
\begin{small}
    \begin{equation}
        {\rm log}\,(\frac{L_{\rm 0.5-2\,keV}^{\rm gas}} {\rm erg\,s^{-1}})_{\rm Center} =1.82(\pm0.08)\,{\rm log}\,\frac{L_{\rm IR}} {L_\odot} +22.26(\pm0.75),
    \label{equation4}
    \end{equation}
\end{small}
\begin{small}
    \begin{equation}
        {\rm log}\,(\frac{L_{\rm 0.5-2\,keV}^{\rm gas}} {\rm erg\,s^{-1}})_{\rm Disk} =0.88(\pm0.19)\,{\rm log}\,\frac{L_{\rm IR}} {L_\odot} +30.71(\pm1.66).
    \label{equation5}
    \end{equation}
\end{small}

According to the calibrations, SFR/$L_{\rm IR}$ $\thicksim$ 1.5 $\times$ 10$^{-10}$ $M_\odot$ yr$^{-1}$ per $L_\odot$, of \citet{1998Kennicutt} and \citet{2011Murphy}, the steep relation of equation~(\ref{equation4}) is consistent with the fitting result between {\lx} and SFR of \citet{Zhang2024ApJ}, and the flat relation of equation~(\ref{equation5}) is similar to the scaling relations identified by \citet{2020Kouroumpatzakis}. The discrepancy between the center and the disk suggests the presence of distinct patterns of gas heating in different regions of galaxies. In star-forming galaxies, the energy released by young, massive stars as they age and eventually explode as SNe can prevail over gravity, driving the expansion of a hot, high-pressure bubble into the cold ISM, which is shock-heated and swept up \citep{2022Nardini_hxga}. At the center of a galaxy, large amounts of gaseous material rapidly fill these bubbles and are reheated, while hot gas generated in the spiral arms escapes more easily into interarm regions and subsequently cools through radiative processes. In Section~\ref{SN}, we will discuss the energy sources responsible for heating gas in different regions of M51 and explain the physical mechanisms underlying the twofold correlation depicted in Figure~\ref{disk_center}.

\begin{figure*}
    \centering 
    \includegraphics[scale=1]{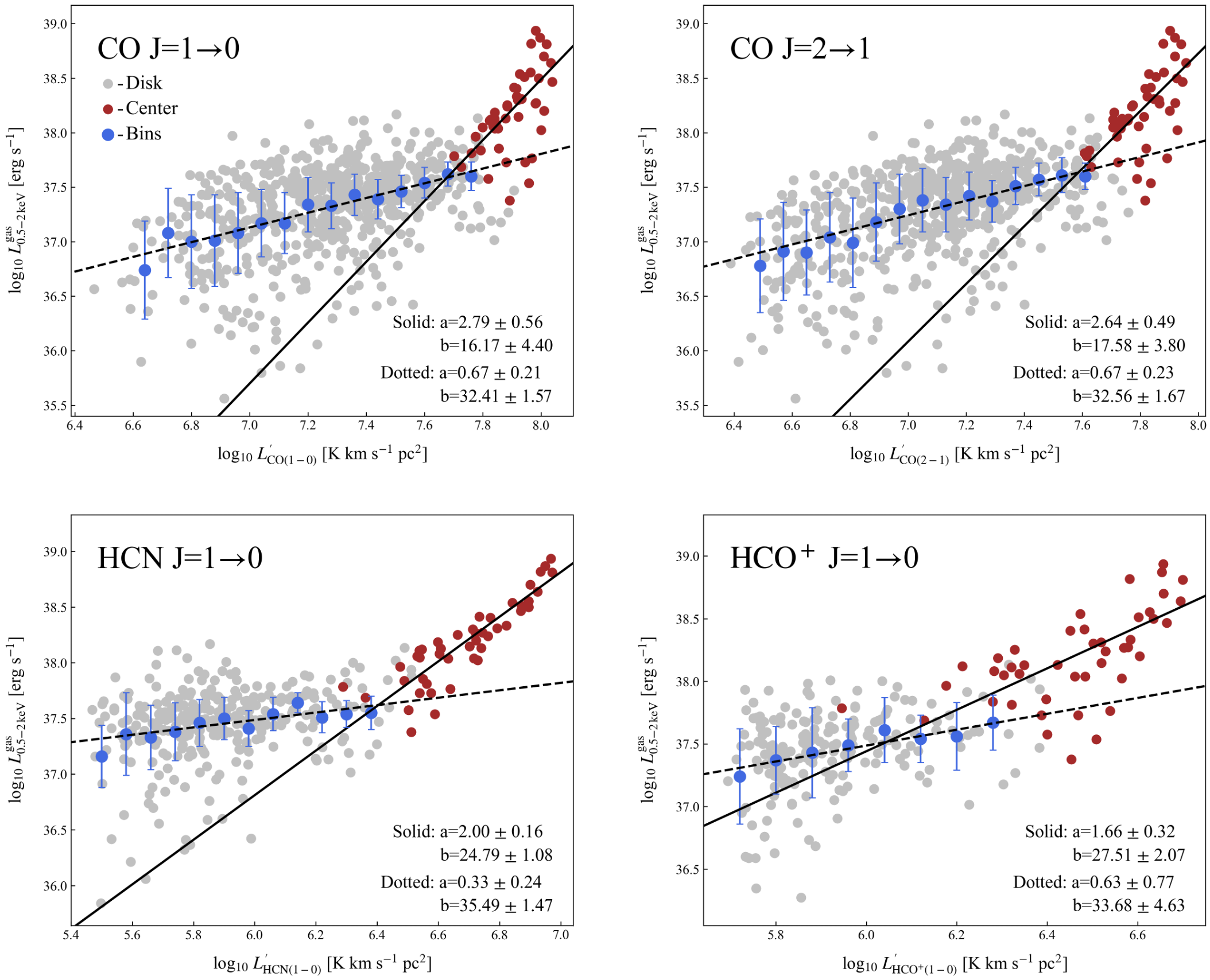}
    \vspace{0.07cm}
    \caption{Correlations between the hot gas emission and the four molecular line luminosities. The symbols (points and lines) are as in Figure~\ref{disk_center}. The parameters a and b of the best fit for ${\rm log}$$L_{\rm 0.5-2\,keV}^{\rm gas}$ = ${\rm a}$ ${\rm log}$$L^\prime_{\rm gas}$ $+$ ${\rm b}$ are shown in the lower right corner of each panel.}
    \vspace{0.3cm}
    \label{4emission_lines}
\end{figure*}

\subsection{Correlations between Hot Gas Luminosity and Cold Gas Molecular Line Emissions}\label{Lx-Lgas}

The cold gas as the raw material for star formation shows a power-law relationship ($\Sigma_{\rm SFR}\propto \Sigma^n_{\rm gas}$, $n\approx 1.4$) with SFR \citep{1998Kennicutt}. Moreover, in denser environments, the dense gas is tightly and linearly correlated with SFR \citep{Gao&Solomon2004ApJ_a,Gao&Solomon2004ApJS_b}. Considering the link mediated by star formation, the hot gas is also expected to have a corresponding correlation with the cold and dense gas. Figure~\ref{4emission_lines} shows the relationships between the luminosity of the hot gas and the four molecular line emissions serving as gas tracers. Similar to the total infrared band, two rising trends are presented in the four emission lines.

We assume a linear model ${\rm log}$$L_{\rm 0.5-2\,keV}^{\rm gas}$ = ${\rm a}$ ${\rm log}$$L^\prime_{\rm gas}$ $+$ ${\rm b}$ for the {\lx}${-}$$L^\prime_{\rm gas}$ relations and the best parameters of the linear regressions are shown in Figure~\ref{4emission_lines}. The highly consistent results between {\coone} and {\cotwo} indicate relatively stable CO(2-1)-to-CO(1-0) line ratios in the center and the disk of M51, in agreement with previous studies \citep[e.g.,][]{2021Brok_MNRAS,2013Leroy_AJ,2021Yajima_PASJ}. For {\hcnone}, it shows the least scatter for the {\lx}${-}$$L^\prime_{\rm gas}$ relations in the center. However, the luminosities of {\hcop} are substantially weaker than those of HCN within the inner $\thicksim$ 3 kpc of M51. In the center, it is observed that the scatter in the four {\lx}${-}$$L^\prime_{\rm gas}$ relations tends to increase in regions with lower {\lx} values. These relatively dim regions of hot gas luminosity are mainly located at the outskirts of the center, where the environment is closer to the galactic disk than that of the innermost regions. This suggests that the increasing scatter is associated with the environmental variation. 

\begin{figure*}
    \centering 
    \includegraphics[scale=1]{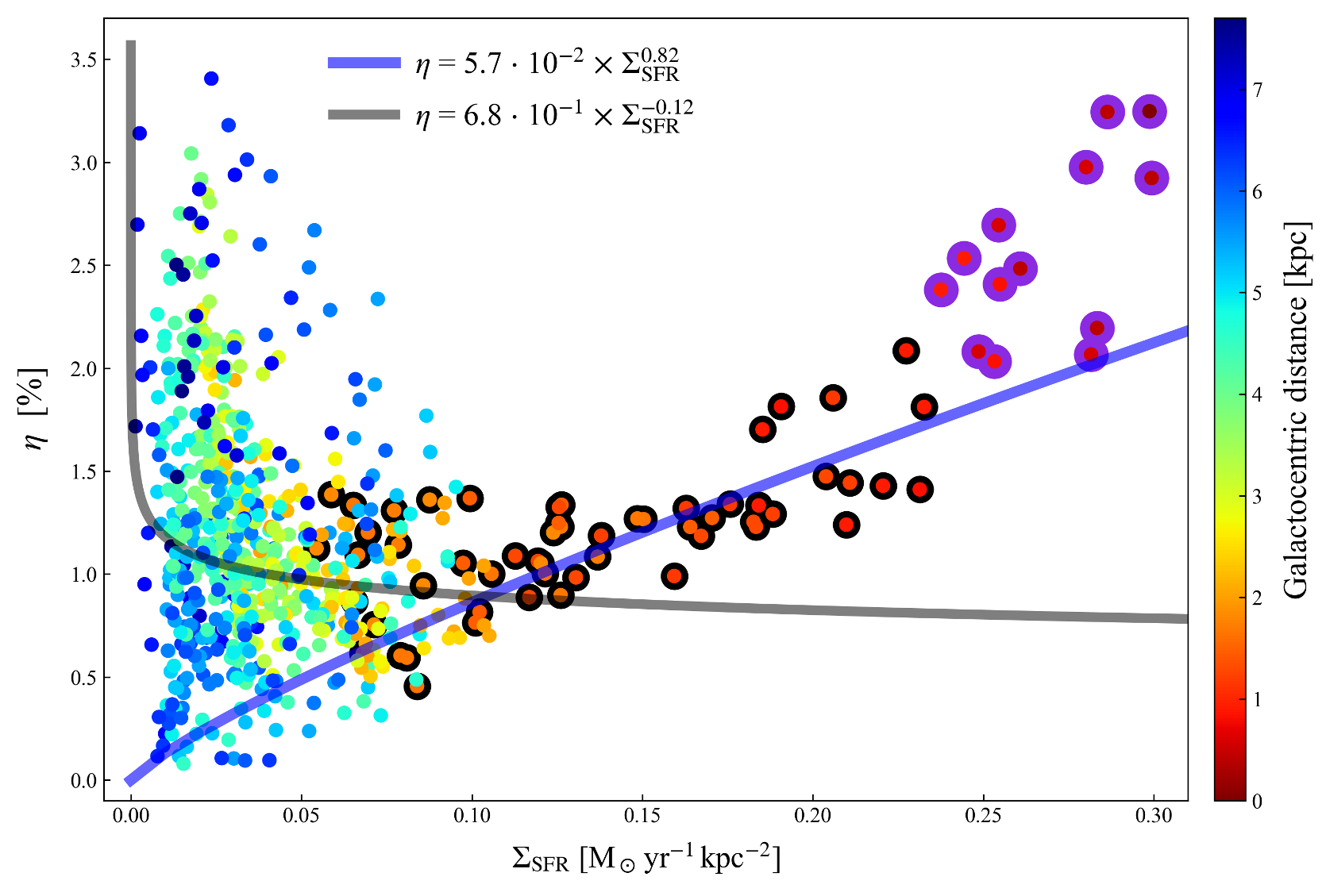}
    \vspace{0.07cm}
    \caption{Correlations between X-ray radiation efficiency and star formation rate surface density at the kiloparsec scale. The color code represents the distance from the galactic center. The purple- and black-bordered points correspond to regions within 0.6 kpc and 2 kpc of M51, respectively. The blue and gray curves are the derived relationships between $\eta$ and $\Sigma_{\rm SFR}$ for the center and the disk.} 
    \vspace{0.3cm}
    \label{SNII}
\end{figure*}

\section{Discussion} \label{sec:discussion}

\subsection{Contamination by Faint Compact Sources and AGN}\label{AGN}

Several types of intrinsically faint X-ray sources are present in star-forming environments and are difficult to detect with current instruments. These include low-mass X-ray binaries (LMXBs), coronally active binaries (ABs), cataclysmic variables (CVs), and young faint objects (such as protostars, young stellar objects, and young stars). Although intrinsically faint, their collective radiation can contribute to the diffuse X-ray emission. To determine the impact of this collective contribution on the X-ray emission, we used the \emph{K}-band image to estimate the luminosities of LMXBs, ABs, and CVs according to the scaling relations of \citet{2011Boroson}:
\begin{small}
    \begin{eqnarray}
        &L_{\mathrm{X}}(\mathrm{ABs+CVs})/L_{K}=4.4_{-0.9}^{+1.5} \times 10^{27} \, \mathrm{erg \, s^{-1}} L_{\mathrm{K\odot}}^{\, \, -1}, \\
        &L_{\mathrm{X}}(\mathrm{LMXBs})/L_{K}=10^{29.0 \, \pm \, 0.176} \, \mathrm{erg \, s^{-1}} L_{\mathrm{K\odot}}^{\, \, -1}, 
    \end{eqnarray}
\end{small} 
where $L_{\mathrm{K\odot}}$ is in solar luminosity. We find that the collective emission from ABs and CVs is negligible relative to the diffuse emission, and the contribution from LMXBs is even less significant than that of ABs and CVs. The young faint objects typically exhibit hard X-ray spectra, corresponding to temperatures of a few keV. These values are substantially higher than the sub-keV temperatures, which are characteristic of the hot gas in star-forming galaxies. \citet{2011BogdanMNRAS} calibrated the scaling relation between the young faint objects and SFR in the 2${-}$10 keV energy band, obtaining $L_{\rm X}$/SFR $\thicksim$ 1.7 $\times$ 10$^{38}$ erg s$^{-1}$ per $M_\odot$ yr$^{-1}$. Using this calibration, we find that the hard X-ray emission from the young faint objects in most regions is only a few percent of the diffuse X-ray luminosity, indicating that the contribution of these faint compact sources can be safely ignored. 

M51 hosts a low-luminosity active galactic nucleus (AGN). In the infrared band, the total infrared of galaxies correlates well with far-infrared compared to near- and mid-infrared and is therefore less contaminated by AGN emission \citep{Nardini2008MNRAS, Juneau2009ApJ, U2012ApJS}. For X-ray emission, the absorption-corrected luminosity (0.5${-}$4 keV) of the AGN is 4.5 $\times$ 10$^{38}$ ergs s$^{-1}$ \citep{Terashima2001ApJ}, which accounts for only 10\% of the diffuse luminosity of the central 33$\arcsec$ region. In this work, although we have excluded the Seyfert 2 nucleus, the AGN effects still need to be considered, as its feedback extends out to a distance of $\thicksim$ 500 pc \citep{Querejeta2016A&A}. Given that the shock velocity of the extranuclear cloud is $\thicksim$ 700 km s$^{-1}$ \citep{Terashima2001ApJ}, we estimate that the nuclear activity took place approximately 0.7 $\times$ 10$^6$ yr ago. Based on the gas temperature $kT$ = 0.76 keV, the emission measure EM = $n_e^2\,V$ = 1.3 $\times$ 10$^{62}$ cm$^{-3}$ of the central 33$\arcsec$ region, and using the expressions in \citet{Richings2010ApJ}, we derive a cooling time $\tau_c$ $\approx$ 5.9 $\times$ 10$^7$ yr and a mass $M_{\rm hot}$ $\approx$ 1.9 $\times$ 10$^6$ $M_{\odot}$ for the hot gas, where $n_e$ is the electron number density and $V$ is the volume. These derived values are consistent with the results reported by \citet{Palumbo1985ApJ}. The long cooling time indicates that the past nuclear activity certainly contributes to the observed X-ray emission. However, we find that the contribution of the low-luminosity AGN is less than 50\% of that of the core-collapse SNe in the center of M51 and has a small influence on the {\lx}${-}$$L_{\rm IR}$ relations. The detailed discussion of energy source of the diffuse X-ray emission is in Section~\ref{SN}.

For the molecular emission lines, theoretical models predict an enhancement of the HCN/{\hcop} abundance ratio in X-ray-dominated regions \citep{Lepp1996A&A, Meijerink2007A&A}. Observations also provide evidence of HCN enhancement in nearby galaxies hosting AGNs \citep{2003Kohno,2012Davies,2016Izumi}. In Figure~\ref{4emission_lines}, the enhanced HCN line luminosities compared to {\hcop} are consistent with the theoretical predictions and the previous observations. This enhancement can be attributed to the increased HCN abundance, driven by infrared pumping via the AGN \citep{2015Matsushita_ApJ,2023Stuber_A&A}. The high abundance electron surrounding the AGN may also contribute to the enhanced HCN emission \citep{2017Goldsmith_ApJ} and disrupt the formation of {\hcop} \citep{2014Zhang}. Using a standard CO-to-H$_2$ conversion factor of $\alpha$$_{\rm CO}$ = 4.3 $M_{\odot}$ pc$^{-2}$ (K km s$^{-1}$)$^{-1}$ \citep{Bolatto2013ARA&A}, we derive a molecular gas mass of $M_{\rm mol}$ $\approx$ 2.7 $\times$ 10$^9$ $M_{\odot}$ for the central 33$\arcsec$ region. In conclusion, unlike the negligible contamination in the total infrared and the soft X-ray bands, the AGN of M51 has a significant effect on the HCN and {\hcop} emissions.

\subsection{X-Ray Radiation Efficiency on the Kiloparsec Scale}\label{SN}

The hot gas is powerd by the intense winds from massive stars and SNe and it is associated with regions of active star formation \citep[see][]{2004Tyler_ApJ,2019Fabbiano_cxro.book}. If we assume that the core-collapse SNe are the primary sources of energy heating the ISM, with $E_\mathrm{SN}$ = 10$^{51}$ erg of mechanical energy released in a single explosion \citep{Grimes2005ApJ, Bogd2008MNRAS, 2012Mineo2}, the relations between the hot gas luminosity and the total infrared band can allow us to estimate the X-ray radiation efficiency ($\eta$ $\equiv$ {\lx}/$\dot{E}_\mathrm{SN}$, where the $\dot{E}_\mathrm{SN}$ is the SN mechanical energy injection rate) of SNe on the kiloparsec scale \citep{Li2013MNRAS2}. Considering the explosion rate for core-collapse SNe is $\thicksim$ 1 SN per 100 yr per 1 $M_{\odot}$/yr \citep{Botticella2012A&A}, the rate of SN mechanical energy injection is $\dot{E}_\mathrm{SN}$ $\approx$ 3.2 $\cdot$ 10$^{41}$ $\times$ SFR erg/s. Due to the weak effect of the AGN on the total infrared luminosity, it can be converted to star formation rate surface density using the prescription of \citet{2011Murphy}: 
\begin{equation}
    \left(\frac{\Sigma_{\rm SFR}}{M_\odot\ {\rm yr}^{-1}\ {\rm kpc}^{-2}}\right) = 1.48\times 10^{-10} \left(\frac{\Sigma_{\rm IR}}{L_\odot\ {\rm kpc}^{-2}} \right),
    \label{equation8}
\end{equation} 
By combining equations (\ref{equation4}), (\ref{equation5}), (\ref{equation8}) and $\dot{E}_\mathrm{SN}$ $\approx$ 3.2 $\cdot$ 10$^{41}$ $\times$ SFR erg/s, we obtain relations between the X-ray radiation efficiency and the star formation rate surface density for the center and the disk of M51:
\begin{eqnarray}
    &\eta=5.7 \cdot 10^{-2} \times \Sigma_{\rm SFR}^{0.82}, \quad  (\rm Center) 
    \label{equation9} \\
    &\eta=6.8 \cdot 10^{-1} \times \Sigma_{\rm SFR}^{-0.12}, \quad (\rm Disk) 
    \label{equation10} 
\end{eqnarray}

These relations indicate that X-ray radiation efficiency is correlated with star formation activity in different regions of galaxies. However, studies of nearby galaxies show no significant correlation of $\eta$ with SFR at the global scale \citep[e.g.,][]{2012Mineo2,Li2013MNRAS1,Li2013MNRAS2}. Simulations of large-scale outflows in star-forming galaxies also find that the energy loading factor of hot outflows, $\eta_{E,\rm{hot}}$ (the ratio of the outgoing energy of hot outflows to SNe-injected energy), is insensitive to $\Sigma_{\rm SFR}$ (e.g., $\eta_{E,\rm{hot}}$ $\thicksim$ $\Sigma_{\rm SFR}^{0.14}$) \citep{Kim2020ApJ}. To explore whether the X-ray radiation efficiency is associated with $\Sigma_{\rm SFR}$ on the kiloparsec scale, we plot $\eta$ and $\Sigma_{\rm SFR}$ for each region in Figure \ref{SNII}. Within the inner 2 kpc of M51, the X-ray radiation efficiency and $\Sigma_{\rm SFR}$ agree well with our derived equation (\ref{equation9}). This demonstrates that the core-collapse SNe are indeed primary sources of energy powering the ISM in the central range. We infer that the X-ray radiation efficiency increasing with the $\Sigma_{\rm SFR}$ is the essential reason for the steep relation of {\lx} with total infrared luminosity or star formation rate.

Within the inner 0.6 kpc region, the X-ray radiation efficiencies are slightly higher than the prediction from equation (\ref{equation9}). A possible explanation is related to the X-ray luminosity excess for $\eta$ $\equiv$ {\lx}/$\dot{E}_\mathrm{SN}$. Since the majority of clusters in the center of M51 are young and massive \citep{Scheepmaker2009A&A, Reina-Campos2017MNRAS, Turner2021MNRAS}, the energy injection from the strong stellar winds generated by these massive stars becomes non-negligible for powering the hot gas emission \citep{Fierlinger2016MNRAS}. In addition, the past nuclear activity also contributes to the observed X-ray luminosities due to the long cooling timescale of the hot gas (see Section~\ref{AGN}). Comparing the X-ray radiation efficiency with the predicted one by equation (\ref{equation9}), we estimate that the collective contribution of stellar winds and nuclear activity is less than 50\% of the energy released by core-collapse SNe. However, in the disk of M51, the X-ray radiation efficiencies of numerous regions deviate significantly from the values predicted by equation (\ref{equation10}). This suggests that the diffuse X-ray emission in the disk is not dominated by gas heated by core-collapse SNe. Type Ia SNe are expected to play an important role in heating gas since the old stellar clusters ($\textgreater$ 200 Myr) of the galaxy are mainly distributed in this region \citep{Scheepmaker2009A&A, Raskin2009ApJ}.

\section{Summary}\label{sec:summary}

In this Letter, we analyze scaling relations between {\lx}, $L_{\rm IR}$, and $L^\prime_{\rm gas}$ across the entire galaxy resolved at a kiloparsec scale and explore the physical origin of these relations. A sharp decrease is present in the hot gas surface brightness profile within the inner region of 2 kpc radius. The inner $r$ $\lesssim$ 2 kpc region and the outer region beyond 2 kpc are defined as the center and the disk of M51, respectively. We find that the soft X-ray emission of the hot gas presents a markedly extended structure compared to the total infrared band. However, the ratios between $L_{\rm IR}$ and {\lx} in the central regions are significantly lower than the average $L_{\rm IR}$/{\lx} ratio for M51 and show an increasing trend with the galactic radius, indicating a higher concentration of hot gas toward the center.

On the kiloparsec scale, we find that the hot gas luminosity has different relations with respect to the total infrared luminosity for the center and the disk, respectively. Using the Bayesian method LINMIX\_ERR, we obtain a steep relation with a slope of 1.82 for the center and a flat relation for the average X-ray luminosities of the disk. The similar twofold correlations are also shown in the {\lx}${-}$$L^\prime_{\rm gas}$ relations for the four molecular line emissions. The luminosities of {\hcop} present the largest scatter with the hot gas emission in the center compared to other molecular lines. For HCN, the enhanced luminosities compared to {\hcop} are attributed to the influence of the Seyfert 2 AGN of M51. However, we do not find that the low-luminosity AGN and the unresolved compact sources have a significant effect on the hot gas luminosities. 

To explain the twofold correlations, we estimate that the X-ray radiation efficiency will vary with the star formation rate surface density on the kiloparsec scale, where $\eta$ increases with $\Sigma_{\rm SFR}$ in the center but decreases with $\Sigma_{\rm SFR}$ in the disk. The observed $\eta$ and $\Sigma_{\rm SFR}$ in the center agree well with the prediction from the analytical formula we derived, suggesting that the core-collapse SNe are the primary sources of energy heating the ISM in the inner 2 kpc region. We conclude that the positive correlation between $\eta$ and $\Sigma_{\rm SFR}$ results in the steep {\lx}${-}$$L_{\rm IR}$ and {\lx}${-}$$L^\prime_{\rm gas}$ relations in the central regions of galaxies.


We thank the anonymous referee for very constructive comments and suggestions that significantly improved our work. C.Y.Z. thanks Dr. Hao Chen, Yang Gao, Yu-Heng Zhang, Hui Li, and Shui-Nai Zhang for their helpful suggestions. J.W. acknowledges the National Key R$\&$D Program of China (grant No.2023YFA1607904) and the National Natural Science Foundation of China (NSFC) grants 12033004, 12221003, and 12333002 and the science research grant from CMS-CSST-2021-A06 and the CNSA program D050102. T.C. acknowledges the China Postdoctoral Science Foundation (grant No. 2023M742929).
This research made use of \emph{Chandra} archival data and software provided by the \emph{Chandra} X-ray Center (CXC) in the application package CIAO. This research has made use of SAOImage DS9, developed by Smithsonian Astrophysical Observatory. The IRAM is supported by INSU/CNRS (France), MPG (Germany) and IGN (Spain). This research has made use of the NASA/IPAC Extragalactic Database (NED) which is operated by the Jet Propulsion Laboratory, California Institute of Technology, under contract with the National Aeronautics and Space Administration.


\bibliography{bibliography}{}
\bibliographystyle{aasjournal}

\end{document}